\documentclass[prl,twocolumn, showpacs]{revtex4}
\usepackage{amsmath}
\usepackage{mathpazo,epsfig,amssymb,amsmath,color,graphicx,rotating,subfloat,booktabs}
\usepackage[version=3]{mhchem}
\usepackage{amsfonts}
\usepackage{amsmath}
\usepackage{amssymb}
\usepackage{graphicx}
\usepackage{subfigure}
\usepackage{subfloat}

\begin{document}

\title{Temperature Coefficient of Resistivity in Amorphous Semiconductors}
\author{Ming-Liang Zhang and David A. Drabold}
\affiliation{Department of Physics and Astronomy, Ohio University, Athens, Ohio 45701}
\date{November 14, 2011}

\begin{abstract}
By invoking the microscopic response method in conjunction with a
reasonable set of approximations, we obtain new explicit expressions for the
electrical conductivity and temperature coefficient of resistivity (TCR) in
amorphous semiconductors, especially a-Si:H and a-Ge:H. The predicted TCR
for n-doped a-Si:H and a-Ge:H is in agreement with experiments. The
conductivity from the transitions from a localized state to an extended
state (LE) is comparable to that from the transitions between two localized
states (LL). This resolves a long-standing anomaly, a ``kink" in the experimental $\log _{10}\sigma $
vs. T$^{-1}$ curve.
\end{abstract}

\maketitle

The temperature coefficient of resistivity (TCR) in disordered systems is
an important physical observable, is difficult to compute, and is of technological interest for
microbolometer materials for thermal imaging applications\cite{motda,str}.
Boltzmann or master equations are often used to calculate transport
coefficients in crystalline semiconductors and semi-metals. The low carrier
concentration in these materials results in a low kinetic energy of
carriers. Thus the Landau-Peierls condition: the basic criterion for a
kinetic approach [$\hbar /\tau <\max \{E_{F},k_{B}T\}$] cannot be satisfied%
\cite{pei}, where $\tau $ is the time interval between two collisions
(carrier with disorder and/or phonon), and $E_{F}$ is the Fermi energy of a
material. Then, neither the elastic scattering by disorder nor the inelastic
scattering by a phonon has a well-defined transition probability per unit
time. The situation for amorphous semiconductors (AS) is even more
difficult. Due to the strong electron-phonon (e-ph) interaction for
localized states, any transition involving localized state(s) requires a
reorganization of the vibrational configuration\cite{epjb,short,pss}. Energy
conservation between the initial and final electronic states\cite{pei} for
these intrinsic multi-phonon transitions is violated more seriously than for
the single-phonon processes. Thus the kinetic method is unjustified for AS\cite{pei}.

In the kinetic approach\cite{ma}, it is often supposed that (i) electrical
conduction is fulfilled by the transition from a localized state to another
localized state (LL) and the transition from an extended state to another
extended state (EE)\cite{str,motda}; (ii) the transition from a localized
state to an extended state (LE) and the transition from an extended state to
a localized state (EL) do not directly contribute to conduction; (iii) LE
and EL transitions only maintain the non-equilibrium stationary distribution
of carriers between localized states and extended states during a conduction
process. Although phonon-assisted delocalization\cite{kik,mul} and
photon-excited transient current\cite{her} have been considered intuitively,
rigorous expressions for LE and EL transition contributions to the
conductivity are not yet available.

Because the interaction between the external electromagnetic field and an AS
can be expressed with additional terms in the Hamiltonian, the transport
coefficients can be expressed with transition amplitudes in the Microscopic
Response Method (MRM)\cite{short,pss}. Thus the long time limit required in
the kinetic approach\cite{pei} is avoided. In addition, the MRM categorizes
transport processes with diagrams computed to any given order of residual
interactions\cite{pss}. We have seen that even to zero order in the residual
interactions, LE and EL transitions contribute to conductivity\cite{pss}.
Indeed, if one calculates the electrical conductivity of an AS from the full
density matrix rather than its diagonal elements (master equation),\ one
sees that LE and EL transitions contribute directly to electrical
conduction. Since the MRM\ is equivalent to the density matrix method of Kubo%
\cite{eqv}, the two methods reach the same conclusion.

Disorder scattering in EE transitions driven by field has been treated in the
coherent potential approximation. The conductivity from the EE transitions
depends weakly on temperature\cite{but}, and is the same order of magnitude
as that from LL transitions above room temperature\cite{motda,str}. In this
Letter we apply MRM\cite{short,pss} to derive the contributions to
conductivity from LL, LE and EL transitions solely drive by an external
field. Two examples, the conductivity and TCR of a-Si:H and a-Ge:H are
described.

An accurate conductivity calculation requires (i) the eigenvalues and
eigenvectors of single-electron states and (ii) the eigenfrequencies and
eigenvectors for each normal mode\cite{pss}. To express the conductivity in
terms of accessible material parameters, we approximate the vibrations of an
AS by a continuous medium. Although translational invariance is destroyed in
AS, the standing wave modes are still well-defined. Because most amorphous
materials are isotropic\cite{str,motda} and only acoustic modes are
important for the e-ph interaction in semiconductors\cite{han}, one can use $%
\omega _{\mathbf{k}}=\overline{c}k$ for the vibrational spectrum, where $%
\omega _{\mathbf{k}}$ is the angular frequency for the mode characterized by
wave vector $\mathbf{k}$. $\overline{c}$ is the average speed of sound
defined by $3/\overline{c}^{3}=2/c_{t}^{3}+1/c_{l}^{3}$, where $c_{t}$ and $%
c_{l}$ are the speeds of transverse and longitudinal sound waves, determined
by the bulk and shear modulus. The Debye cutoff wave vector $k_{D}=(6\pi
^{2}n_{a})^{1/3}$ is determined by the number density $n_{a}$ of atoms in an
AS\cite{han}. Because the displacements of atoms satisfy a wave equation,
the transformation matrix $\Delta $ between the vibrational displacements $%
\mathbf{u}_{\mathbf{R}}$ at point $\mathbf{R}$ and the normal coordinates $%
\Theta _{\mathbf{k}}$ characterized by wave vector $\mathbf{k}$ is:
\begin{equation}
\Delta _{\mathbf{Rk}}=(2\pi )^{-3}Ve^{i\mathbf{k}\cdot \mathbf{R}},
\label{k2r}
\end{equation}%
where $V$ is the volume of a sample. For a-Si\cite{str}, $\overline{c}%
=6.21\times 10^{3}$m/s and $k_{D}=1.44\times 10^{10}$m$^{-1}$.

For simplicity, we take localized states to be spherically symmetric\cite%
{nev}. The difference between localized states is expressed by a
single-parameter localization length $\xi _{A}$\cite{motda}. We will use
letter $A$ with or without a natural number subscript to denote a localized
state. For a localized state $A$, denote $\mathbf{R}_{A}$ as the position
vector of the center, the normalized wave function is

\begin{equation}
\phi _{A}(\mathbf{r}-\mathbf{R}_{A})=\pi ^{-1/2}\xi _{A}^{-3/2}e^{-|\mathbf{r%
}-\mathbf{r}_{A}|/\xi _{A}},  \label{a15}
\end{equation}%
where $\mathbf{r}$ is the coordinate of the electron and localization length%
\cite{nev}. Following Mott, $\xi _{A}$ is determined by the eigenvalue $E$
of localized state $\phi _{A}$\cite{nev}: $\xi _{E}=(cZe^{2}/4\pi \epsilon
_{0}\varepsilon )(E_{c}-E)^{-1},$ where $Z$ is the effective nuclear charge
of an atom core, $\varepsilon $ is the static dielectric constant. $E_{c}$
is the mobility edge of the conduction band, and $c$ is a dimensionless
constant. We will focus on n-doped material: transport in the conduction
band, p-doped material may be treated analogously. For a-Si\cite{str}, $%
\varepsilon =11.68$, $Z=4$ and the values of $E_{c}$ are rather dispersed%
\cite{vis,jj,ora}: 0.2-2eV: we will take\cite{jj} $E_{c}=0.5$eV. The most
localized states are dangling bonds, they have the shortest possible $\xi
_{\min }=2.35/2$\AA\ (one half of a bond length). $E=0$ for a dangling bond,
then $c=0.121$. For many AS\cite{aljishi}, in the range of conduction band
tail, the density of localized states (DOS) satisfies
\begin{equation}
N(E)=(n_{loc}/U)e^{-(E_{c}-E)/U},  \label{udos}
\end{equation}%
where $U$ is the Urbach energy, $n_{loc}$ is the total number of localized
states per unit volume. For a-Si, $U\thickapprox 50$meV\cite{ora,weh}, the
number\ density $n_{loc}$ of localized conduction states is\cite{ting} $%
n_{loc}=5/(10.86$\AA $)^{3}$. The exponential $N(E)$ implies that most
localized states in a-Si have a localization length in the range 6-12\AA .
Denote $n$ as the carrier concentration, the Fermi energy $E_{F}$ of a
weakly doped AS is:
\begin{equation}
E_{F}=E_{c}+U\ln (n/2n_{loc}),  \label{fer}
\end{equation}%
When $n\leq 2n_{loc}$, all occupied states are localized at T$=0$K. Ansatz (%
\ref{udos}) only characterizes the band tail states. To describe dangling
bonds, one can (i) introduce a reasonable DOS, e.g. a rectangle or a
Gaussian; (ii) correspondingly modify $E_{F}$ and energy zero-point for
extended states; (iii) add the contribution from the dangling bonds to Eqs.(%
\ref{cle},\ref{ddc}) in the summation(s) over localized states. In this
paper we ignore the small contribution of dangling bonds to the conductivity.

In an AS, an extended state is a packet of Bloch waves of its reference
crystal\cite{vky,scm}, and is labeled by the wave vector of its principal
Bloch wave, or more roughly by the momentum $\mathbf{p}$ of a plane wave\cite%
{but,motda}. For an AS, for which the reference crystal does not exist, a
wave packet constructed from plane waves is still a reasonable approximation
for an extended state. We will use letter $B$ with or without a natural
number subscript to denote an extended state. Excepting EE transitions
driven by an external field, we may approximate an extended state $\chi
_{B_{1}}(\mathbf{r})$ by a plane wave with certain momentum $\mathbf{p}$,
and its eigenenergy is that of the plane wave:
\begin{equation}
\chi _{B_{1}}=V^{-1/2}e^{i\mathbf{p}\cdot \mathbf{r}/\hbar },\text{ \ \ }%
E_{B_{1}}=p^{2}/2m,  \label{ted}
\end{equation}%
where the zero-point of energy for extended states is at the mobility edge $%
E_{c}$. The attraction between an electron and an atom core may be
approximated by a screened Coulomb potential\cite{han}. For a-Si:H\cite{str}%
, we approximate its Thomas-Fermi wave vector by the value\cite{han} for
c-Si $q_{TF}=1.7$\AA $^{-1}$.

With the foregoing approximations, the velocity matrix elements in the
expressions of conductivity can be computed\cite{4tsf}. One can also obtain
the static displacements of the atoms in a localized state induced by the
e-ph interaction and the reorganization energy\cite{epjb} for transitions
involving localized state(s)\cite{4tsf}, which are essential input for the
conductivity.


\begin{figure}[th]
\centering
\par
\subfigure[]{\includegraphics[scale=0.15]{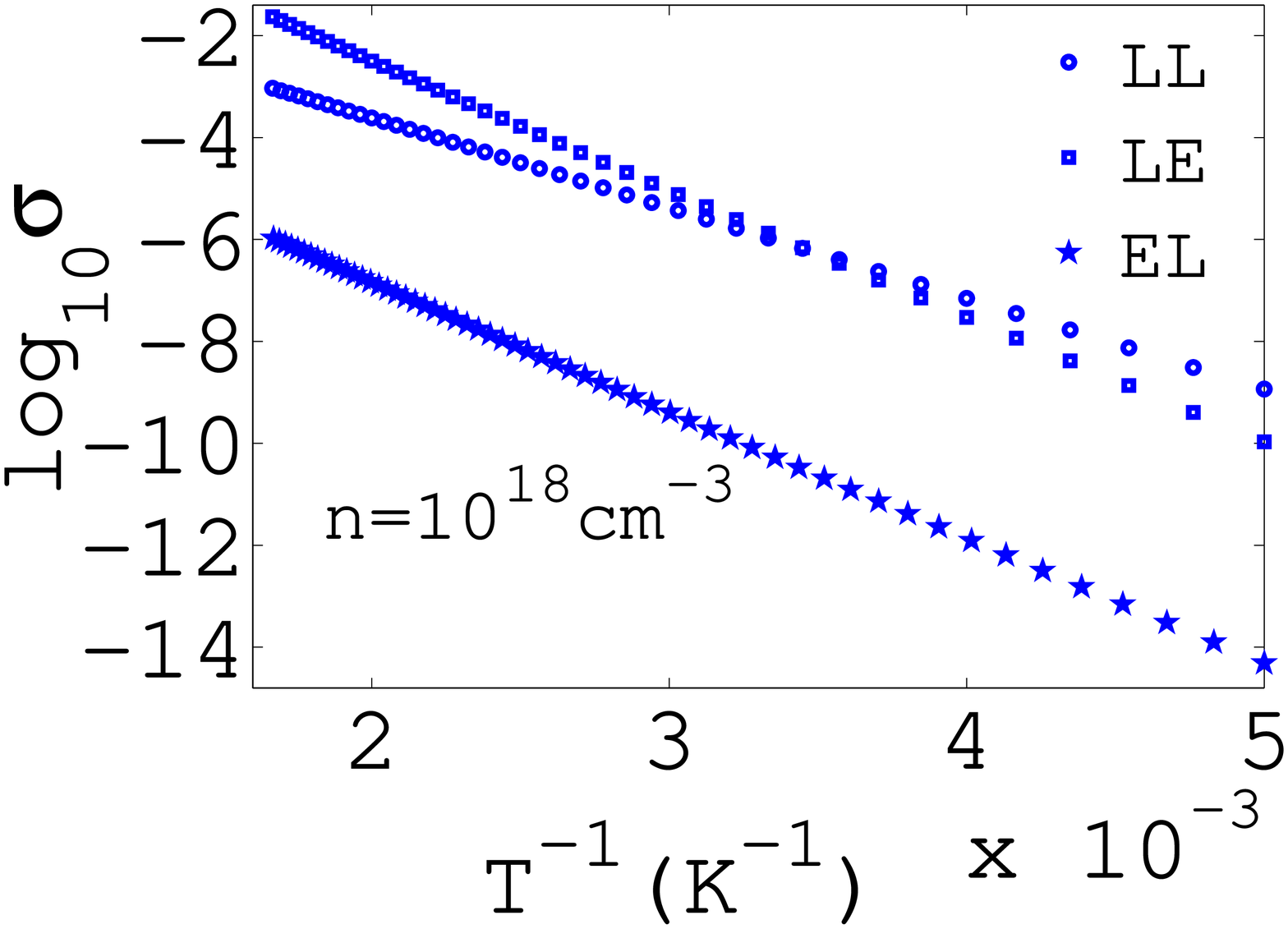}\label{18con}}\hfill %
\subfigure[]{\includegraphics[scale=0.15]{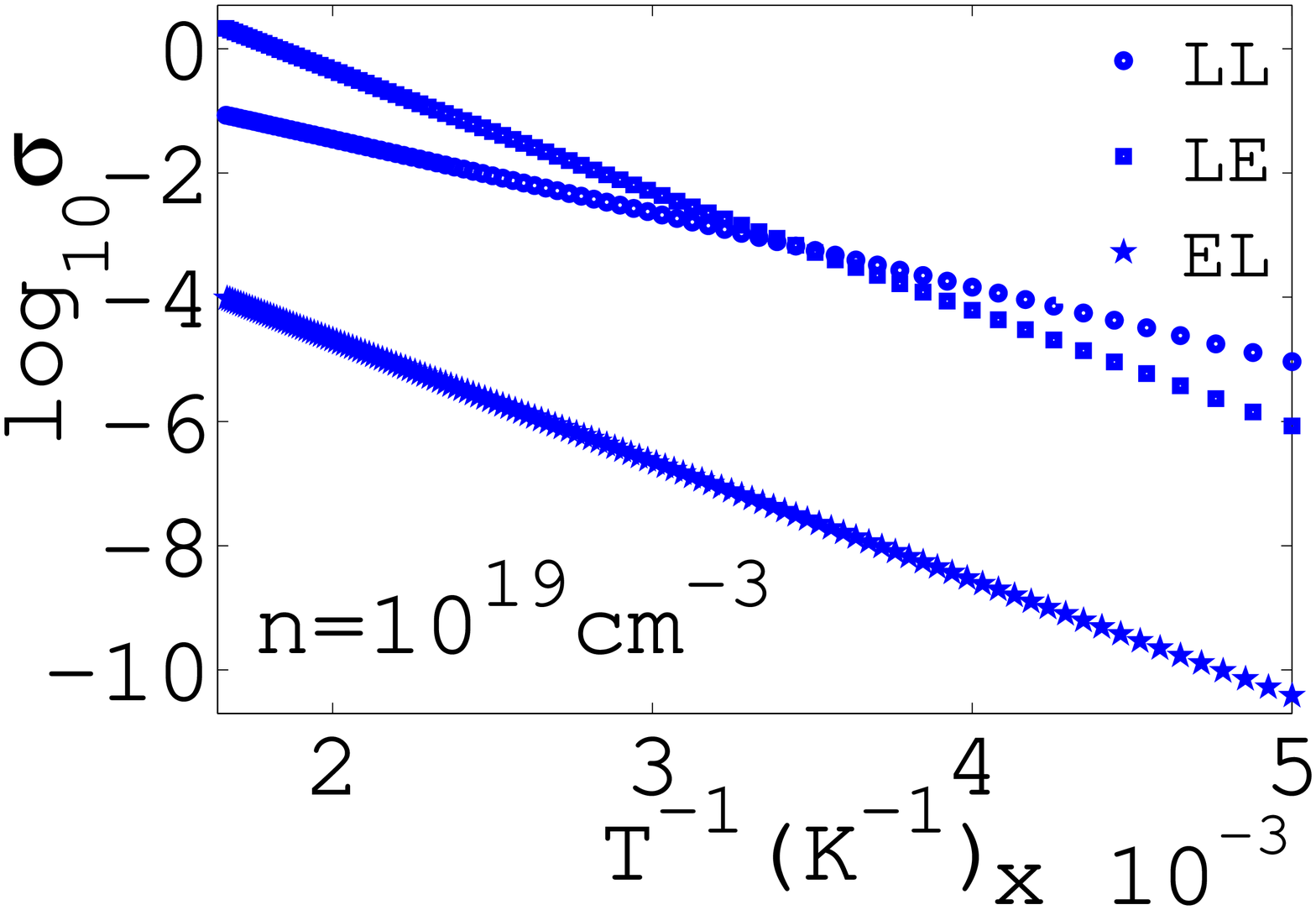}\label{19con}}\hfill %
\caption{(color online) $\log _{10}\protect\sigma $ vs. $1/T$ in two n-doped a-Si:H samples
at $\protect\omega =0$ with carrier concentration $10^{18}$ and $10^{19}$cm$%
^{-3}$. The unit of $\sigma $ is ohm$^{-1}$cm$^{-1}$.}
\label{con}
\end{figure}

We first calculate the conductivity from the LE transitions (line 2b of
Table 4 in Ref.\cite{pss}). When $k_{B}T\geq \hbar \overline{\omega }$ (the
first peak of vibrational spectrum, $\hbar \overline{\omega }=$ 232K for a-Si%
\cite{kam}), the two time integrals I$_{B_{1}A\pm }$ can be approximated by
an asymptotic expansion\cite{4tsf}, and the vibrational degrees of freedom
are integrated out.\ For LE transitions, we first sum over final electronic
states $\sum_{B}$ for a fixed localized state $A$. It is convenient to use a
spherical coordinate system with $\mathbf{R}_{A}$ as the origin and the wave
vector direction $\mathbf{k/|k|}$ of the incident electromagnetic wave as
polar axis (z axis)\cite{4tsf}. Because AS are isotropic, the angular part
of the $\mathbf{p}$ integral can be carried out. One can show that $\sigma
_{xy}=\sigma _{yx}=0$ and $\sigma _{xx}=\sigma _{yy}=\sigma $. Because (i)
the center $\mathbf{R}_{A}$ of $\phi _{A}$ must be a neighbor of the
observation point of current density, and (ii) the factors in the
conductivity do not depend on $\mathbf{R}_{A}$, $\sum_{A}\rightarrow \frac{%
4\pi }{3}\overline{\xi }^{3}\int_{-\infty }^{E_{c}}dEN(E)$, where $\overline{%
\xi }=cZe^{2}/(4\pi \epsilon _{0}\varepsilon U)$ is the most probable
localization length. For a-Si, $\overline{\xi }=11.75$\AA , is quite close
to the experimental value\cite{qg,stut} 10\AA . The conductivity from LE
transitions is\cite{4tsf}:%
\begin{equation*}
\left\{
\begin{array}{c}
Re \\
Im%
\end{array}%
\right. \sigma (\omega )=\frac{cZe^{2}(n_{loc}\frac{4\pi }{3}\overline{\xi }%
^{3})}{4\pi \epsilon _{0}\varepsilon U}\frac{8ne^{2}}{3\pi \hbar ^{3}m^{2}}%
\int_{0}^{\infty }d\xi f(E_{A})
\end{equation*}%
\begin{equation}
\int_{0}^{\infty }dp[1-f(E_{B_{1}})]\frac{p^{4}}{(E_{B_{1}}^{0}-E_{A}^{0})}%
\frac{\xi \exp (-\frac{cZe^{2}}{4\pi \epsilon _{0}\varepsilon U\xi })}{%
(1+p^{2}\xi ^{2}/\hbar ^{2})^{4}}  \label{cle}
\end{equation}%
\begin{equation*}
\frac{\sqrt{\pi }\hbar }{2(k_{B}T\lambda _{BA})^{1/2}}[e^{-\frac{\lambda
_{BA}}{4k_{B}T}(1+\frac{\hbar \omega _{BA}-\hbar \omega }{\lambda _{BA}}%
)^{2}}\pm e^{-\frac{\lambda _{BA}}{4k_{B}T}(1+\frac{\hbar \omega _{BA}+\hbar
\omega }{\lambda _{BA}})^{2}}],
\end{equation*}%
where $\theta _{\alpha }^{A}=(M_{\alpha }\omega _{\alpha }/\hbar
)^{1/2}\Theta _{\alpha }^{A}$\cite{epjb} and $\hbar \omega _{BA}=E_{B}-E_{A}$%
. $\lambda _{BA}=\frac{1}{2}\sum_{\alpha }\hbar \omega _{\alpha }(\theta
_{\alpha }^{A})^{2}$ is the reorganization energy for transition $\phi
_{A}\rightarrow \chi _{B}$\cite{epjb}. One can show that $\lambda _{BA}$
decreases with $\xi _{A}$.

Next we consider the conductivity from EL transitions driven by a field
(line 6a of Table 5 in Ref.\cite{pss}). Because the field-matter coupling is
Hermitian, the conductivity for EL transition driven by a field may be
obtained from Eq.(\ref{cle}) by exchanging $\phi _{A}$ and $\chi _{B}$ and
noticing that $\lambda _{AB}=\lambda _{BA}$. For the LE transition driven by
transfer integral and the EL transition driven by e-ph interaction, one does
not have this symmetry\cite{epjb,pss}.

To obtain the conductivity from LL transitions (line 2a of Table 4 in Ref.%
\cite{pss}), the velocity matrix elements $w_{\parallel }^{AA_{1}}$ and $%
v_{\parallel }^{A_{1}A}$ are computed with approximation (\ref{a15}) in a
spherical coordinate system with $\mathbf{R}_{A}$ as origin and $\mathbf{R}=%
\mathbf{R}_{A_{1}}-\mathbf{R}_{A}$ as polar axis\cite{4tsf}; they
exponentially decay with $R=|\mathbf{R}|$. $\sum_{AA_{1}}$ can be carried
out by\cite{4tsf} first considering a fixed $\phi _{A}$ and scanning $\phi
_{A_{1}}$ at all possible $R$ with different $\xi _{A_{1}}$. The
conductivity from LL transition driven by field becomes\cite{4tsf}%
\begin{equation*}
\left\{
\begin{array}{c}
Re \\
Im%
\end{array}%
\right. \sigma (\omega )=\frac{4\pi }{3}\overline{\xi }^{3}[\frac{%
cZe^{2}n_{loc}}{4\pi \epsilon _{0}\varepsilon U}]^{2}\int_{0}^{\infty }\frac{%
d\xi _{1}}{\xi _{1}^{2}}\exp (-\frac{cZe^{2}}{4\pi \epsilon _{0}\varepsilon
U\xi _{1}})
\end{equation*}%
\begin{equation*}
\int_{0}^{\infty }\frac{d\xi _{2}}{\xi _{2}^{2}}\exp (-\frac{cZe^{2}}{4\pi
\epsilon _{0}\varepsilon U\xi _{2}})[1-f(E_{A_{1}})]f(E_{A})
\end{equation*}%
\begin{equation*}
\frac{\sqrt{\pi }\hbar }{2(\lambda _{A_{1}A}k_{B}T)^{1/2}}[e^{-\frac{\lambda
_{A_{1}A}}{4k_{B}T}(1+\frac{\hbar \omega _{A_{1}A}-\hbar \omega }{\lambda
_{A_{1}A}})^{2}}\pm e^{-\frac{\lambda _{A_{1}A}}{4k_{B}T}(1+\frac{\hbar
\omega _{A_{1}A}+\hbar \omega }{\lambda _{A_{1}A}})^{2}}]
\end{equation*}%
\begin{equation}
\int_{0}^{R_{c}}R^{2}dR\frac{4\pi }{3}ne^{2}\frac{(w_{\parallel
}^{AA_{1}}-v_{\parallel }^{A_{1}A})(v_{\parallel }^{A_{1}A})^{\ast }}{%
2(E_{A}^{0}-E_{A_{1}}^{0})},  \label{ddc}
\end{equation}%
where $R_{c}$ is the radius of physical infinitesimal volume elements\cite%
{pss}. $\lambda _{A_{1}A}=\frac{1}{2}\sum_{\alpha }\hbar \omega _{\alpha
}(\theta _{\alpha }^{A_{1}}-\theta _{\alpha }^{A})^{2}$ is the
reorganization energy for transition $\phi _{A}\rightarrow \phi _{A_{1}}$.
To simplify notation, we used $\xi _{2}$ instead of $\xi _{A_{1}}$, used $%
\xi _{1}$ instead of $\xi _{A}$.

\begin{figure}[th]
\centering
\par
\subfigure[]{\includegraphics[scale=0.15]{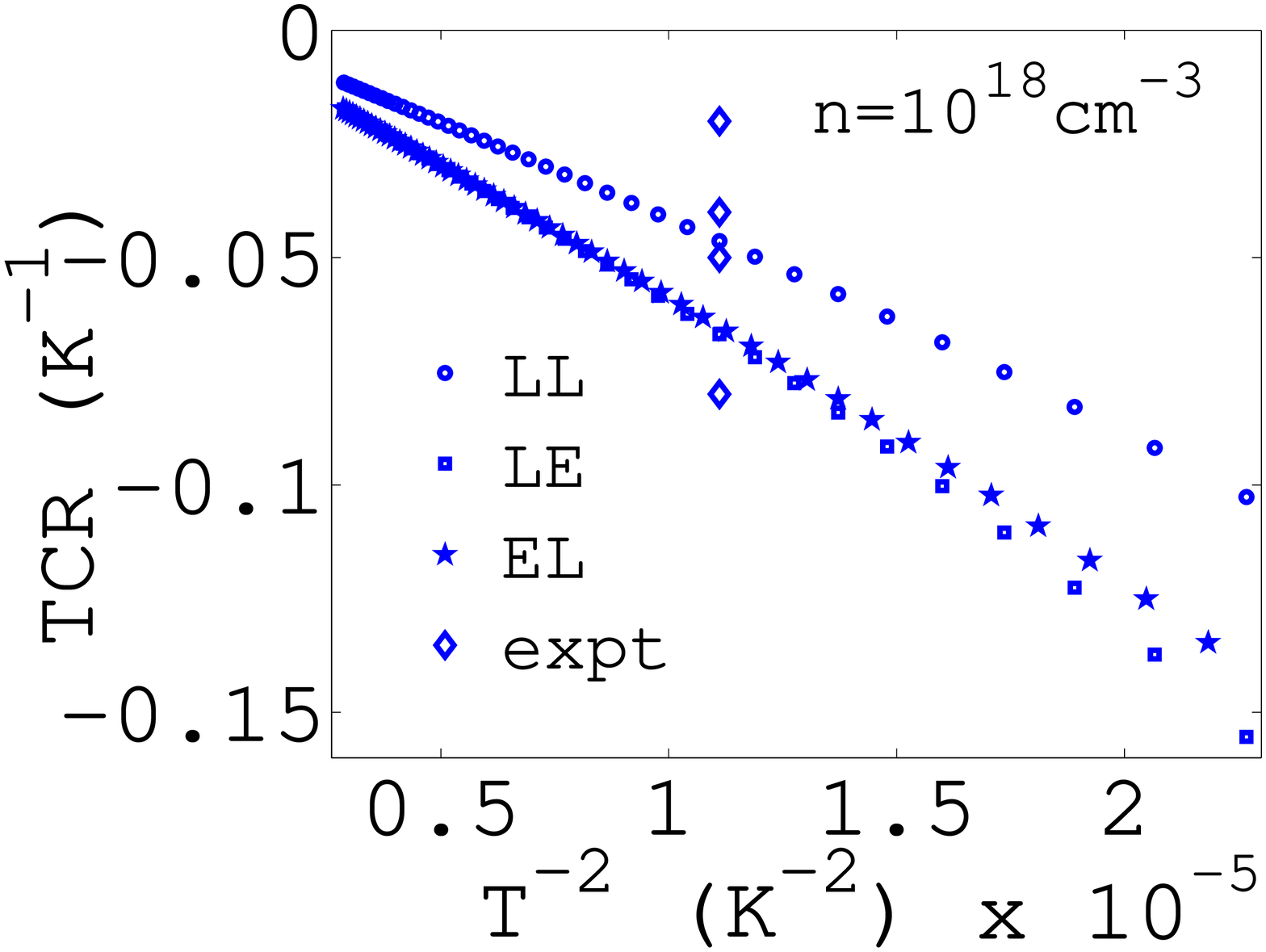}\label{18tcr}}\hfill %
\subfigure[]{\includegraphics[scale=0.15]{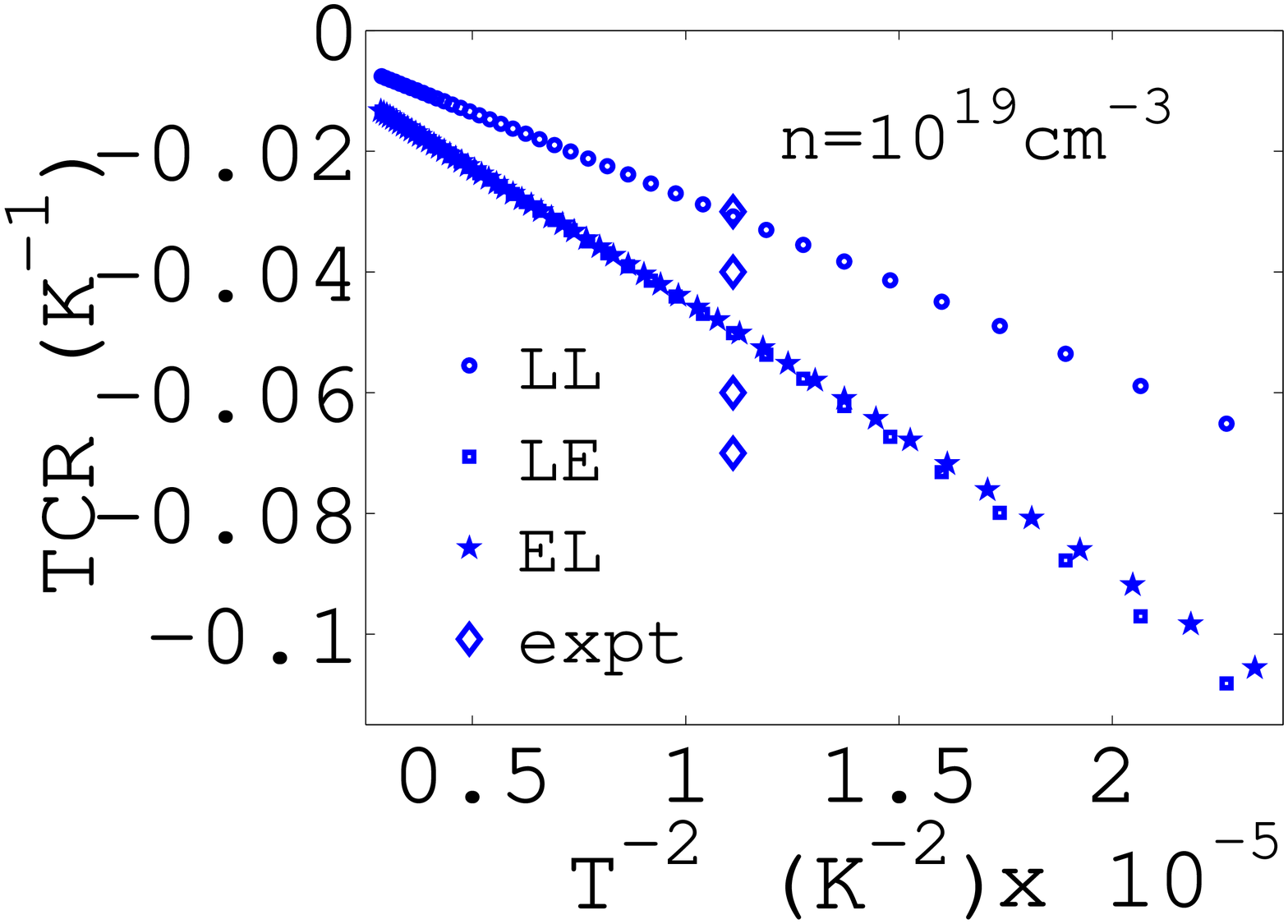}\label{19tcr}}\hfill %
\caption{(color online) TCR vs. $T^{-2}$ in two n-doped a-Si:H samples at $\protect\omega %
=0 $ Hz with carrier concentration $10^{18}$ and $10^{19}$cm$^{-3}$,
experimental values are taken from Refs.\protect\cite{tcr,sai,gar}.}
\label{tcr}
\end{figure}

Eqs.(\ref{cle},\ref{ddc}) express conductivity and its associated
temperature dependence upon several material parameters: $n$, $n_{loc}$, $U$%
, $E_{c}$, $\varepsilon $, $q_{TF}$ and $\overline{c}$. Re$\sigma $
as a function of temperature T is plotted in Fig.\ref{con} for two n-doped
a-Si:H samples\cite{str,bey} with carrier concentration $n=10^{18}$ and $%
10^{19}$cm$^{-3}$ [the unit of $\sigma $ is ohm$^{-1}$cm$^{-1}$]. The
LE conductivity is the same order of magnitude as that from LL
transitions, while the contribution from EL transitions is $10^{-4}-10^{-3}$
of that from LL\ transitions. Thus the calculated conductivity is the same
order magnitude as the observed ones, cf. Fig.8 of Ref.\cite{sai}. From Eqs.(%
\ref{cle},\ref{ddc}), one can easily compute the temperature coefficient of
resistivity (TCR) $\varkappa =\rho ^{-1}\frac{d\rho }{dT}=-\sigma ^{-1}\frac{%
d\sigma }{dT}$. The corresponding TCR vs. T$^{-2}$ is plotted in Fig.\ref%
{tcr}. If there are several processes contributing to conductivity in a
material, according to MRM, the total conductivity $\sigma $ of the material
is $\sigma =\sum_{j}\sigma _{j}$, where $\sigma _{j}$ is the conductivity
from the j$^{th}$ process (LL, LE and EL ect.)\cite{pss}. The overall TCR $%
\varkappa $ relates to the TCR\ for each process by $\varkappa =\sigma
^{-1}\sum_{j}\sigma _{j}\varkappa _{j}$, where $\varkappa _{j}=-\sigma
_{j}^{-1}d\sigma _{j}/dT$.\ At 300-350K, the observed TCR is in range -0.02
to -0.08 for a n-doped a-Si:H with $n=10^{18}$cm$^{-3}$\cite{tcr,sai,gar}.
The calculated TCR from LL transitions is smaller than the experimental
data, this is solid evidence that the contribution from LE transitions is
important.
\begin{figure}[th]
\centering
\par
\subfigure[]{\includegraphics[scale=0.15]{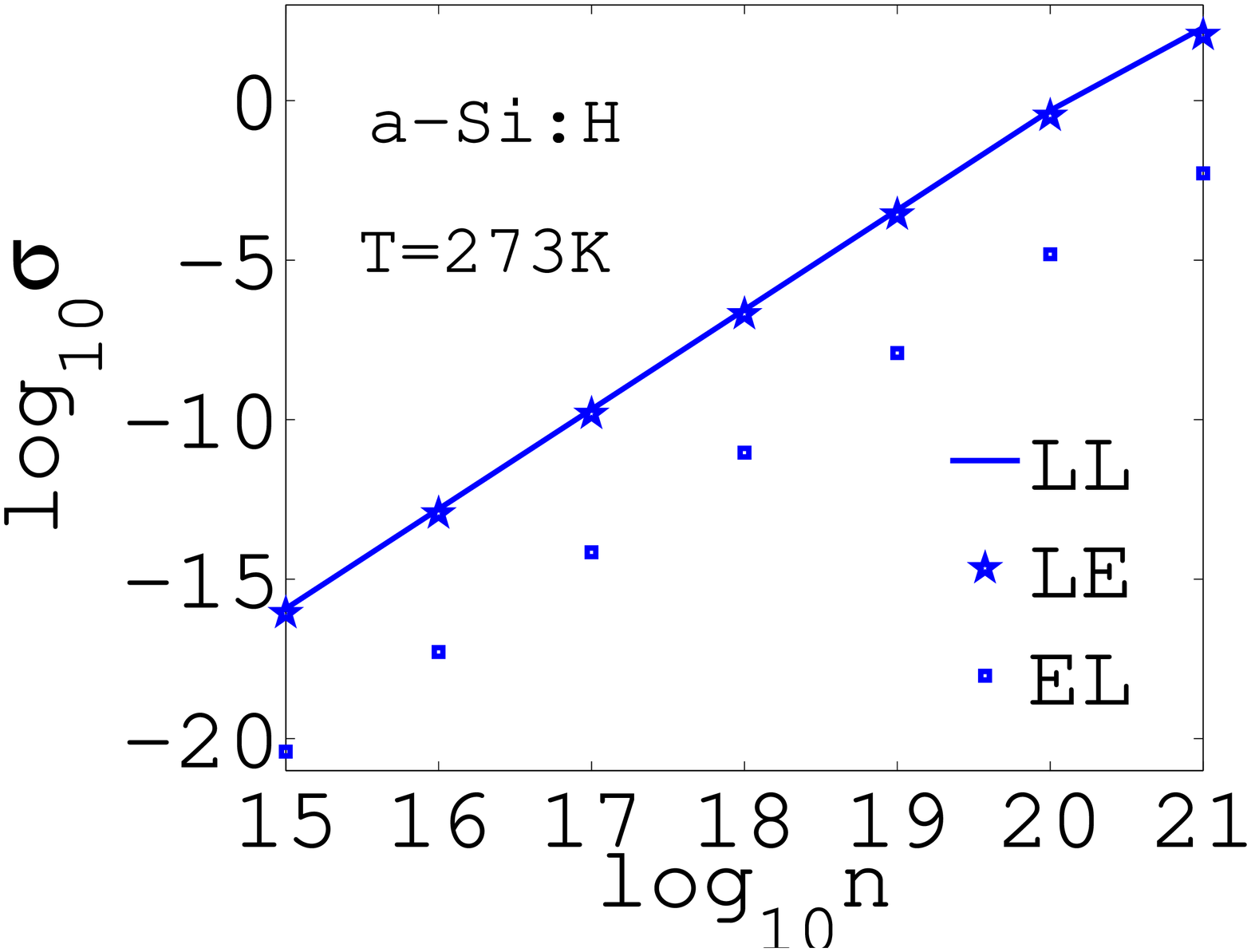}\label{c273}}\hfill %
\subfigure[]{\includegraphics[scale=0.15]{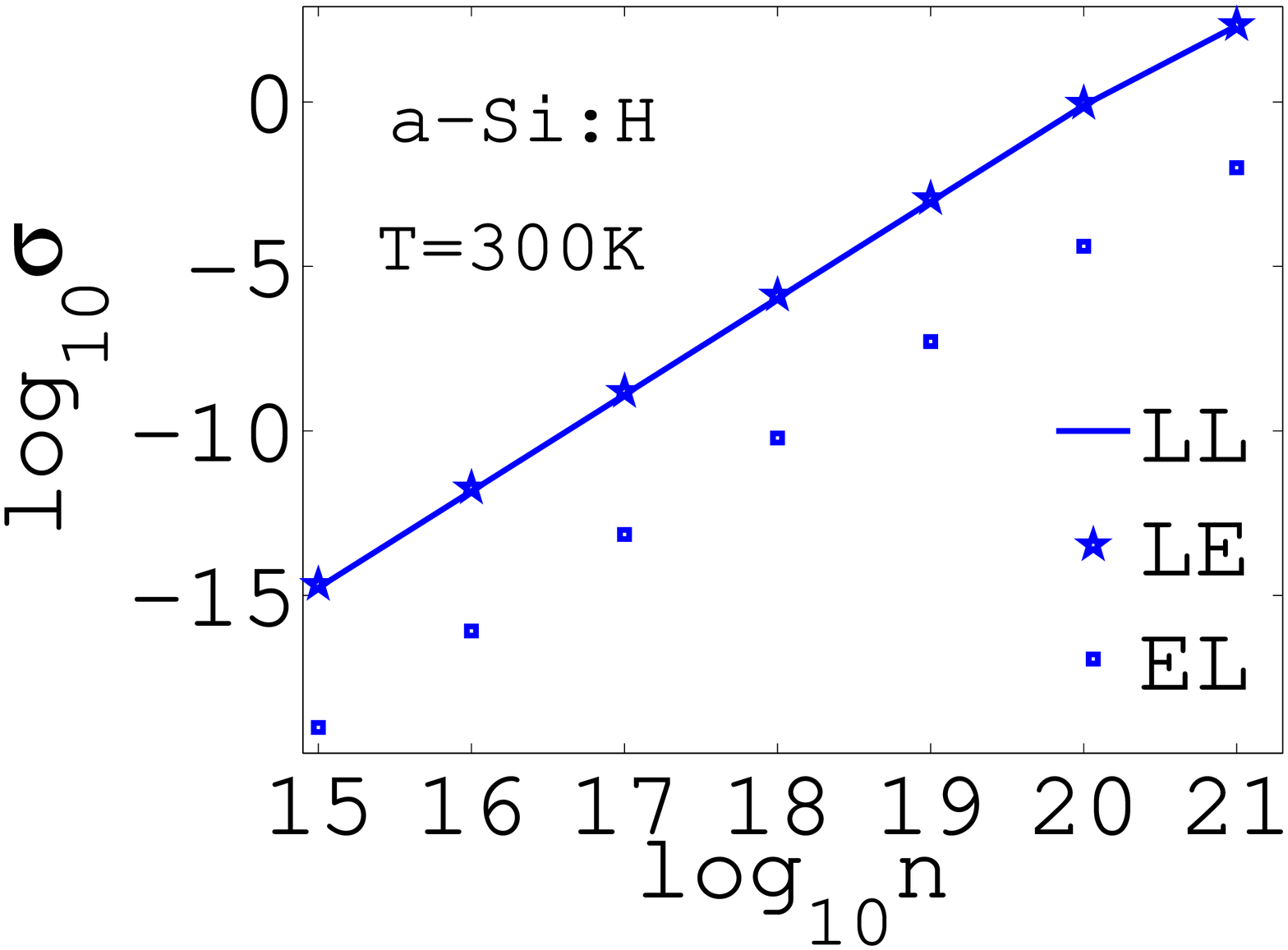}\label{c300}}\hfill %
\subfigure[]{\includegraphics[scale=0.15]{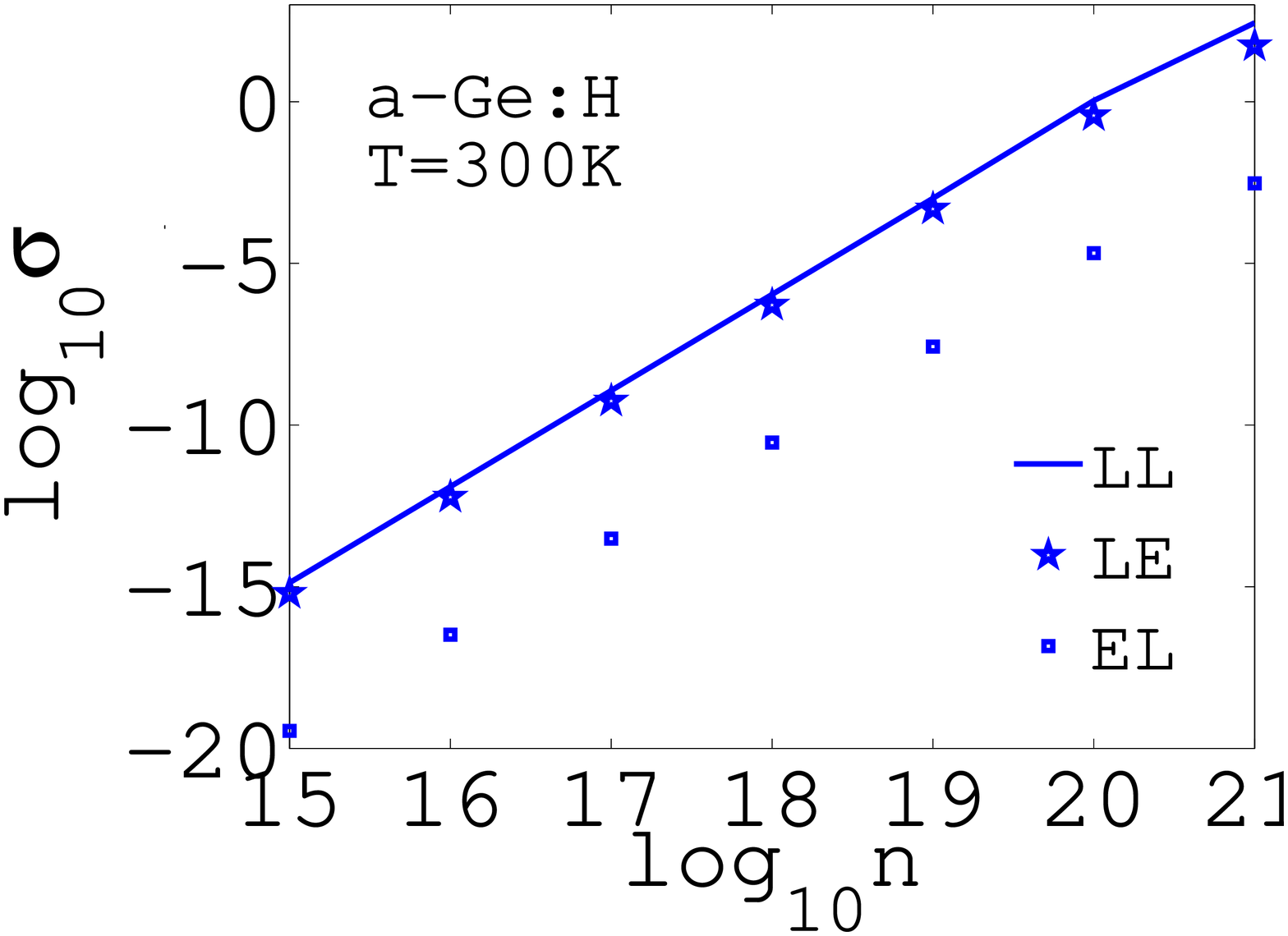}\label{c300Ge}}\hfill %
\subfigure[]{\includegraphics[scale=0.15]{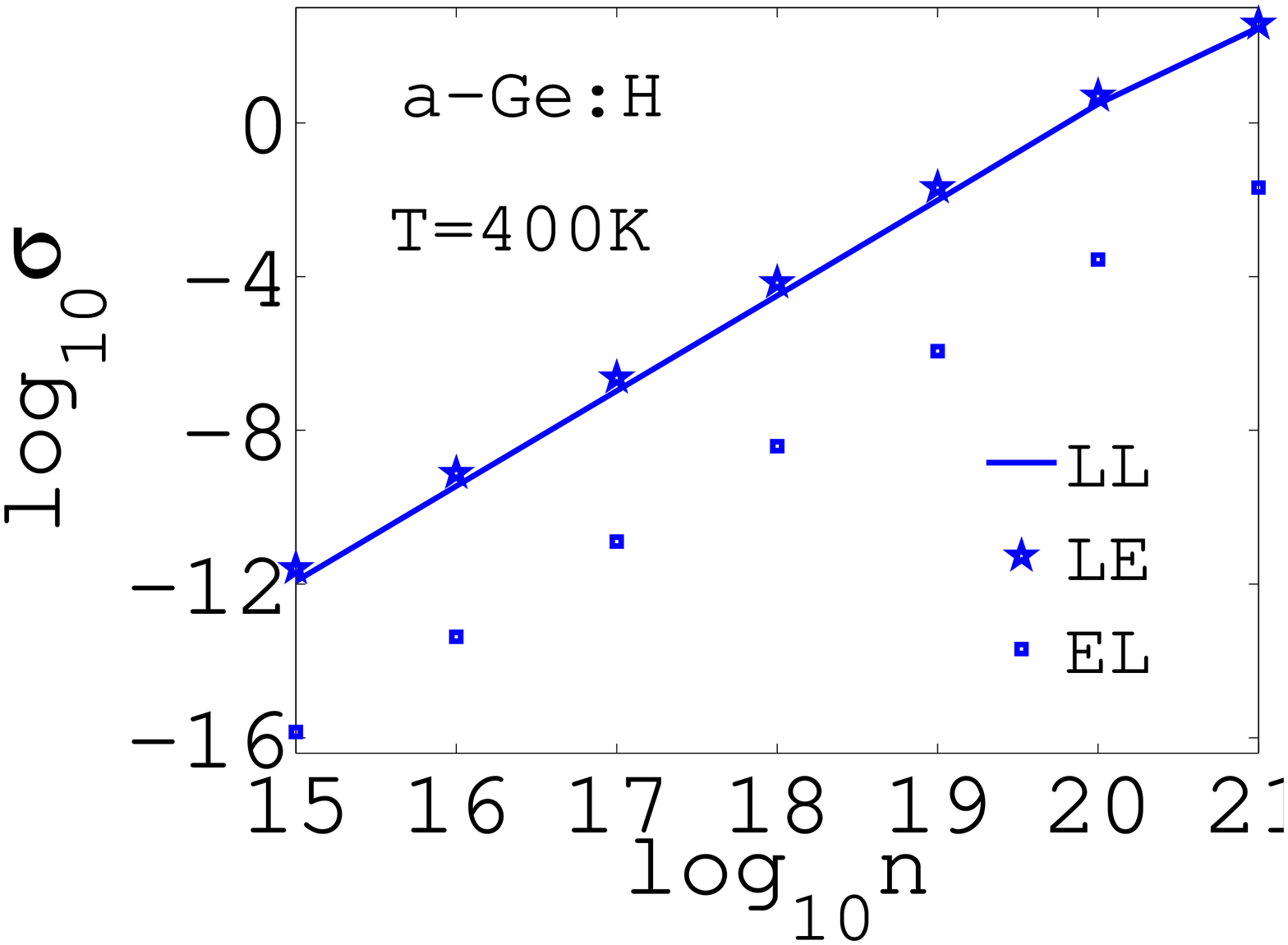}\label{c400Ge}}\hfill
\caption{(color online)$\log _{10}\protect\sigma $ vs. $\log _{10}n$ for n-doped a-Si:H
and a-Ge:H. The unit of $\sigma $ is ohm$^{-1}$cm$^{-1}$, the unit of n is cm$^{-3}$.}
\label{conc}
\end{figure}

In a-Si:H, a long-standing puzzle is (i) there is a kink in the observed $%
\log _{10}\sigma (T)$ vs. 1/T curve; and (ii) for very different doping
concentrations, the kink temperatures collapse into a narrow range\cite{over}.
We resolve these problems. The crossing temperature $T^{\ast }$ of $\sigma
_{LL}(T)$ and $\sigma _{LE}(T)$ is the key to understanding the kink. When $%
T<T^{\ast }$, LL transitions are the main conduction mechanism; for $%
T>T^{\ast }$, LE transitions dominate conductivity. If one forced a single
Arrhenius fit to the overall conductivity, the formal activation energy
would be different below and above T$^{\ast }$. Thus one has a kink in the $\log
_{10}\sigma (T)$ vs. 1/T curve. The non-exponential behavior as indicated in
Eqs.(\ref{cle},\ref{ddc}) also has some role for the observed kink. For LL
and LE transitions, we compute linear fits for the calculated $\log
_{10}\sigma (T)$ vs. T$^{-1}$, for which the norms of the residuals are 0.03
(LL) and 0.15 (LE) for 10$^{18}$cm$^{-3}$; 0.05 (LL) and 0.20 (LE) for 10$%
^{19}$cm$^{-3}$. This is consistent with the deviation from linear relation
in the measured mobility vs. 1/T curve, cf. Fig. 7.10 of \cite{str}.

From Fig.\ref{18con} and \ref{19con}, we can see that T$^{\ast }$ decreases
with $n$: $T^{\ast }=$294K for 10$^{18}$cm$^{-3}$, $T^{\ast }=$286K for 10$%
^{19}$cm$^{-3}$. This is consistent with the trend found in experiments: $%
T^{\ast }=$400K for [PH$_{3}$]/[SiH$_{4}$]=10$^{-6}$ and $T^{\ast }=$333K
for [PH$_{3}$]/[SiH$_{4}$]=10$^{-2}$, cf. Fig. 3.1 of Ref.\cite{over}. Fig. %
\ref{c273} and \ref{c300} plot $\log _{10}\sigma $ vs. $\log _{10}n$ at
T=273K and 300K. We see that at T=273K (300K), the contribution from LL
transitions is larger (smaller) than that from LE transitions for carrier
concentration from 10$^{15}$ to 10$^{21}$cm$^{-3}$. For very different
carrier concentrations (10$^{6}$ times different), the kink temperatures fall
near 273-300K, consistent with fact (ii). In Fig. \ref{c300Ge} and %
\ref{c400Ge}, we plot $\log _{10}\sigma $ vs. $\log _{10}n$ for n-doped
a-Ge:H at T=300K and 400K. $T^{\ast }$s fall between 300-400K, are higher
than those for a-Si:H. It agrees with the observations, compare Fig. 3.1 and
3.6 of\cite{over}.

The compositional atomic orbital and/or their relative phases for the
states close to E$_{F}$ in the valence band (VB) are very different to those
for states close to E$_{F}$ in the conduction band (CB). Because $%
U\varpropto \langle \lbrack (V_{a}-V_{c})/V_{c}]^{2}\rangle _{av}$ and $%
E_{c}\varpropto \langle (V_{a}-V_{c})/V_{c}\rangle _{av}$, where $V_{a}$ and
$V_{c}$ are the potential of an AS and its reference crystal\cite{vky}, $%
\langle \rangle _{av}$ denotes configurational and state average in AS\cite%
{scm}. Larger $U$ and $E_{c}$ means stronger disorder, that implies smaller $%
\xi $ or larger $\lambda $, i.e. smaller $\sigma $ and larger $\varkappa $.

Our approach is not restricted to one component systems with weak electronic
correlation. If one has a reasonable single-electron DOS in which the
correlation between electrons is already taken into account, it is not
difficult to calculate the e-ph coupling in multi-component system, e.g. the
e-ph interaction induced by the optical modes in VO$_{1.83}$. The remaining
procedure is exactly like here.

In conclusion, the microscopic response method expresses transport
coefficients with transition amplitudes\cite{short,pss} rather than
transition probability per unit time, which enables the method to be used
with amorphous semiconductors for which the Landau-Peierls condition is violated%
\cite{pei}. The conductivities from the three simplest transitions: LL, LE
and EL transitions driven by field are expressed by several material
parameters. The conductivity from LE transitions is as important as that
from the LL transitions. The combination is responsible for the kink in the
experimental $\log_{10} \sigma $ vs. $T^{-1}$ curve. The LE transition is
critical in determining the TCR.

This paper provides new analytical form for $\sigma $ and $\varkappa $,
suitable for amorphous semiconductors. A desirable extension would be a full
\textit{ab initio} evaluation of all MRM diagrams using quantities from
density functional theory. This complex task is underway.

\begin{acknowledgements}
We thank for support from the U.S. Army Research
Laboratory and the U. S. Army Research Office under grant number
W911NF-11-1-0358.
\end{acknowledgements}


\begin{thebibliography}{99}
\bibitem{motda} N. F. Mott and E. A. Davis, \textit{Electronic Processes in
Non-crystalline Materials}, Clarendon Press, Oxford, (1971).

\bibitem{str} R. A. Street, \textit{Hydrogenated Amorphous Silicon},
Cambridge Univresity Press, Cambridge (1991).

\bibitem{pei} R. Peierls, \textit{Surprises in Theoretical Physics},
pp121-126, Princeton University Press, Princeton (1979).



\bibitem{epjb} M.-L. Zhang and D. A. Drabold, Eur. Phys. J. B. \textbf{77},
7-23, (2010).

\bibitem{short} M.-L. Zhang and D. A. Drabold, Phys. Rev. Lett. \textbf{105}%
, 186602 (2010).

\bibitem{pss} M.-L. Zhang and D. A. Drabold, Phys. Status Solidi B \textbf{%
248}, 2015-2026, (2011).

\bibitem{ma} A. Miller and E. Abrahams, Phys. Rev. \textbf{120}, 745 (1960).

\bibitem{kik} M. Kikuchi, J. Non-Cryst. Sol. \textbf{59}/\textbf{60}, 25
(1983).

\bibitem{mul} H Muller and P Thomas, J. Phys. C: Solid State Phys. \textbf{17%
}, 5337 (1984).


\bibitem{her} H. Scher, E.W. Montroll, Phys. Rev. B \textbf{12}, 2455 (1975)



\bibitem{eqv} M.-L. Zhang and D. A. Drabold, Phys Rev. E\textbf{83}, 012103
(2011).

\bibitem{but} W. H. Butler, Phys. Rev. B\textbf{31}, 3260, (1985).


\bibitem{han} G. D. Mahan, \textit{Many-Particle Physics}, Second edition,
Plenum Press, New York (1990).










\bibitem{nev} N. F. Mott, \textit{Conduction in Non-Crystalline Materials},
Second edition, Clarendon Press, Oxford (1993).

\bibitem{vis} J. H. Davis, J. Non-Cryst. Solids \textbf{35}, 67-69 (1980).

\bibitem{jj} J. Dong and D. A. Drabold, Phys. Rev. Lett. \textbf{80}, 1928
(1998).

\bibitem{ora} F. Orapunt and S. K. O'Leary, J. Appl. Phys. \textbf{104},
073513 (2008).

\bibitem{aljishi} S. Aljishi, J. D. Cohen, S. Jin and L. Key, Phys. Rev.
Letter \textbf{64}, 2811 (1990).

\bibitem{weh} R. B. Wehrspohn, S. C. Deane, I. D. French, I. G. Gale, M. J.
Powell and R. Br\"{u}ggemann, Applied Physics Letters \textbf{74}, 3374
(1999).

\bibitem{ting} taken from Y.-T. Li and D. A. Drabold's unpublished
calculation on a-Si models with 64 and 216 atoms.

\bibitem{vky} B. Velicky, Phys. Rev. \textbf{184}, 614 (1969).

\bibitem{scm} M.-L. Zhang and D.A. Drabold, Phys. Rev. B\textbf{78}, 195208
(2008).

\bibitem{4tsf} M.-L. Zhang and D.A. Drabold, to be submitted to Phys. Rev.
B.

\bibitem{kam} W. A. Kamitakahara, C. M. Soukoulis and H. R. Shanks, U.
Buchenau and G. S. Grest, Phys. Rev. B \textbf{36}, 6539\ (1987).

\bibitem{qg} Q. Gu, E.A. Schiff, J. Chevrier and B. Equer, Phys. Rev. B%
\textbf{52}, 5695 (1995).

\bibitem{stut} M. Stutzmann and J. Stuke, Solid State Communications,
\textbf{47}, 635-639 (1983).

\bibitem{bey} W. Beyer and H. Mell, in \textit{Amorphous and Liquid
Semiconductors}, p.333, ed. by W. E. Spear, CICL, Edinburgh (1977).

\bibitem{tcr} M. B. Dutt and V. Mittal, J. Appl. Phys. 97, 083704 (2005).




















\bibitem{sai} D. B. Saint John, H.-B. Shin, M.-Y. Lee, S. K. Ajmera, A. J.
Syllaios, E. C. Dickey, T. N. Jackson, and N. J. Podraza, J. Appl. Phys.%
\textbf{110}, 033714 (2011).

\bibitem{over} H. Overhof and P. Thomas, \textit{Electronic Transport in
Hydrogenated Amorphous Semiconductor}, Sec. 3.1, Springer-Verlag, Berlin
(1989).

\bibitem{gar} M. Garcia, R. Ambrosio, A. Torres, A. Kosarev, Journal of
Non-Crystalline Solids 338--340, 744--748 (2004).

\end{thebibliography}
\end{document}